\documentclass[mathpazo]{ata}
\usepackage{graphicx}
\usepackage{float}
\usepackage{color}
\usepackage{subfigure}

\begin{document}
\title{The Mean Time to Absorption on Horizontal Partitioned Sierpinski Gasket Networks}

\author[Zhang et.~al.]{Zhizhuo Zhang\affil{1},
       Bo Wu\affil{2}\comma\corrauth and Zuguo Yu\affil{3}}
\address{  \affilnum{1}\ School of Mathematics, Southeast University, Nanjing 210023, P.R. China\\
				\affilnum{2}\ School of Applied Mathematics, Nanjing University of Finance and Economics, Nanjing 210023, P.R. China\\
               \affilnum{3}\ Key Laboratory of Intelligent Computing and Information Processing of Ministry of Education and Hunan Key Laboratory for Computation and Simulation in Science and Engineering, Xiangtan University, Hunan 411105, P.R. China}
\email{{\tt zhizhuo\_zhang@163.com} (Z. Zhang), {\tt bowu8800@nufe.edu.cn} (B. Wu), {\tt yuzg@xtu.edu.cn} (Z. Yu)}


\begin{abstract}
The random walk is one of the most basic dynamic properties of complex networks, which has gradually become a research hotspot in recent years due to its many applications in actual networks. An important characteristic of the random walk is the mean time to absorption, which plays an extremely important role in the study of topology, dynamics and practical application of complex networks. Analyzing the mean time to absorption on the regular iterative self-similar network models is an important way to explore the influence of self-similarity on the properties of random walks on the network. The existing literatures have proved that even local self-similar structures can greatly affect the properties of random walks on the global network, but they have failed to prove whether these effects are related to the scale of these self-similar structures.
In this article, we construct and study a class of Horizontal Partitioned Sierpinski Gasket network model based on the classic Sierpinski gasket network, which is composed of local self-similar structures, and the scale of these structures will be controlled by the partition coefficient $k$. Then, the analytical expressions and approximate expressions of the mean time to absorption on the network model are obtained, which prove that the size of the self-similar structure in the network will directly restrict the influence of the self-similar structure on the properties of random walks on the network. Finally, we also analyzed the mean time to absorption of different absorption nodes on the network to find the location of the node with the highest absorption efficiency.
\end{abstract}

 \clc{O29} 

\maketitle

\section{Introduction}

Due to its applications in the fields of social sciences, engineering, telecommunication networks and biological networks, complex network science has gradually become a research hotspot in recent years\cite{albert2002statistical,dorogovtsev2002evolution,newman2003structure,boccaletti2006complex}.
The random walk on the network is a research direction in the field of complex networks that has attracted much attention, because it can intuitively describe the dynamics of complex networks \cite{newman2018networks}.
In the field of complex networks, the research of random walk is generally used to detect the community structure in the network\cite{rosvall2008maps}, to segment the network\cite{grady2006random} and to study the corresponding properties of the resistance network\cite{newman2018networks}, and so on.
In addition, the random walk on complex networks has its application value in many practical fields. Current research has shown that the related properties of random walks can be used in the field of communication and information to study a series of issues such as information transmission\cite{chau2011analysis}, data collection\cite{zheng2014data,lee2015towards}, communication quantification and prediction\cite{el2006optimal,liu2012exact}, information latency\cite{chau2011analysis}, communication and search costs\cite{roberto2009low,lin2008dynamic}, and computer vision\cite{gopalakrishnan2009random}; in the field of biology, random walk is used to model and study the spread of infectious diseases and the metabolic flux of organisms\cite{gopalakrishnan2009random,yu2013maximal}.
The most basic problem in the properties of random walk is the first passage time(FPT), which is defined as the number of steps required by the initial node in the network to reach the target node for the first time after random walks\cite{newman2018networks}.
Since there may be many paths between these two nodes and randomness of walking, the first passage time is uncertain. Naturally, the mean first passage time(MFPT) between two nodes has received more attention, which has important application value in studying the transmission cost of wireless networks\cite{li2012random,el2006optimal}.
But the mean first passage time only describes the local information of the network. In order to reveal the global random walk properties of the network, the mean time to absorption(MTA) is further developed on the basis of it, which is defined as the average value of the mean first passage time of all nodes in the network to the absorptive node\cite{chen1999dynamics,meyer2012exact}.
MTA directly reflects the efficiency of other nodes in the network to reach absorptive node through random walks and therefore plays an important role in the selection of data collection nodes and best absorptive sites for exciton transport in polymer and electron transfer on a fractal photosynthetic antenna\cite{peng2014mean,brin1998anatomy,blumen1981energy,van2000exciton,
hwang2012first,hwang2012effective,heijs2004trapping,bar1998mean}.

The analytical expression of MTA on a general random network is difficult to be obtained, but on a regular iterative network with a specific structure, it is possible to find the iterative expression of MTA according to the iterative law of the network structure, which is of great significance for further understanding and studying the influence of network structure on random walk \cite{peng2018moments,peng2016scaling,peng2019exact}.
Sierpinski gasket network is a classic self-similar network model, so many scholars analyzed the influence of self-similarity on the topology and dynamics of the network by studying this network and its extended network model \cite{mandelbrot1983fractal,falconer2004fractal,aguirre2009fractal,rammal1984spectrum,chang2007spanning,
chang2008dimer,guyer1984diffusion,haynes2008global,meyer2012exact,dhar1997distribution,lin2002electronic}.
For example, the MTA and the spectrum problem with absorptive nodes on the second-order Sierpinski gasket have been studied by Kozak et al\cite{kozak2002analytic,bentz2010analytic}; then, Zhang et al. made certain improvements to the classic Sierpinski gasket network, and obtained network models with more specific properties, including: deterministic Sierpinski network(DSN)\cite{zhang2007incompatibility}, random Sierpinski network(RSN)\cite{zhang2008random}, evolutionary Sierpinski networks(ESNs)\cite{guan2009unified} and dual Sierpinski gaskets(DSGs)\cite{wu2011random}, and studied the topology and dynamic properties of these network models. On the basis of previous studies, the average trapping time problem on the third-order Sierpinski gasket network model is solved by us, in which the core method is to combine the probability generating function with the iterative method \cite{wu2020average}. Since these networks are constructed in an iterative manner, they have strict self-similar properties. However, actual networks often have a certain degree of self-similarity but will not meet such strict conditions. Therefore, the Sierpinski gasket network was segmented or spliced which are named the half-Sierpinski gasket (HSG)\cite{wu2020average} network model and the joint Sierpinski gasket (JSG)\cite{zhang2020mean} network model respectively. These network mode are different from the original network model in that they lose the global self-similarity and only retain the self-similarity in the local area. Then, we studied the MTA problems on these two types of networks, and found that even though they lost the global self-similar properties, and the analytical expressions of MTA are no longer the same, they still retain the iterative law very similar to the original network.

Therefore, the above works have shown that even the self-similarity of the local area is enough to affect the global random walk properties of the network, but the local self-similar structure in HSG and JSG still occupies the main part of the network. The new question is: when such a self-similarity structure is sufficiently small relative to the whole network , will the influence of self-similarity on random walk be weakened? In order to answer this question, in this article we will perform a more detailed horizontal segmentation of the Sierpinski gasket network and study the MTA problem on the second half of the network. The network model is named Horizontal Partitioned Sierpinski Gasket.
Although the method used in the JSG network model can solve the MTA problem on the partially incomplete Sierpinski gasket network, it cannot deal with the network model whose number of self-similar modules increases exponentially, and it cannot establish the relationship between the segmentation level and the MTA analytical expression. In this paper, an iterative network mode that can handle the above problem will be proposed. On these network model, we can not only solve the analytical expression of the MTA, but also establish the quantitative relationship between the MTA and the residual degree of the network model (the size of the local self-similar structure). Then the question of whether the influence of self-similarity on random walks is restricted by the size of its self-similarity structure is answered. In addition, based on the Horizontal Partitioned Sierpinski Gasket network model, we can also parameterize the location of some nodes, and then analyze the relationship between the MTA and the location of absorptive nodes.

The subsequent content of the article will be divided into the following three sections: in the second section, the construction method of the Horizontal Partitioned Sierpinski Gasket network model will be introduced systematically; in the third section, the calculation method of MTA will be divided into five parts and displayed in turn; finally in the fourth section, we will summarize the article.

\section{The Horizontal Partitioned Sierpinski Gasket}
In this section, we'll cover the Sierpinski Gasket $SG(g)$ and how to split the network horizontally.
As a classical fractal network model, many properties of Sierpinski Gasket, such as spectrum, resistance distance, average absorption time, etc., have been studied by many scholars. The Sierpinski Gasket can be generated by iteration. In Fig 1, we show the network structure in the initial state when $g=0$, and the network structure after two iterations, namely $SG(1)$ and $SG(2)$. The specific iteration of the Sierpinski Gasket can be referred to the previous research and do not repeat. However, it is worth noting the self-similarity of the network, which is especially important in the later calculation. As shown in Fig.1, the network $SG(g)$ can also be composed of three identical regions, each of which can be denoted as $\Gamma(g-1)$. From the self-similarity of the network, we can know that region $\Gamma(g-1)$ is the $g-1$ generation network $SG(g-1)$. Similarly, using region $\Gamma(g)$, namely the network $SG(g)$, we can also construct the network $SG(g+1)$. Where, the three initial nodes in the initial network are denoted as $A$, $B$ and $C$ and the three nodes generated after first generation are denoted as $D$, $E$ and $F$. In addition, we also label the nodes in the network from top to bottom and from left to right, as shown in Fig.1. The total number of nodes and the total number of edges in $SG(g)$ of generation $g$ network are denoted as $N(g)$ and $E(g)$ respectively, which can be obtained from previous work:
$$
N(g)=\frac32(3^g+1)\quad \textrm{and}\quad E(g)=3^{g+1}.
$$

\begin{figure}[t]
\centering
\includegraphics[scale=0.25]{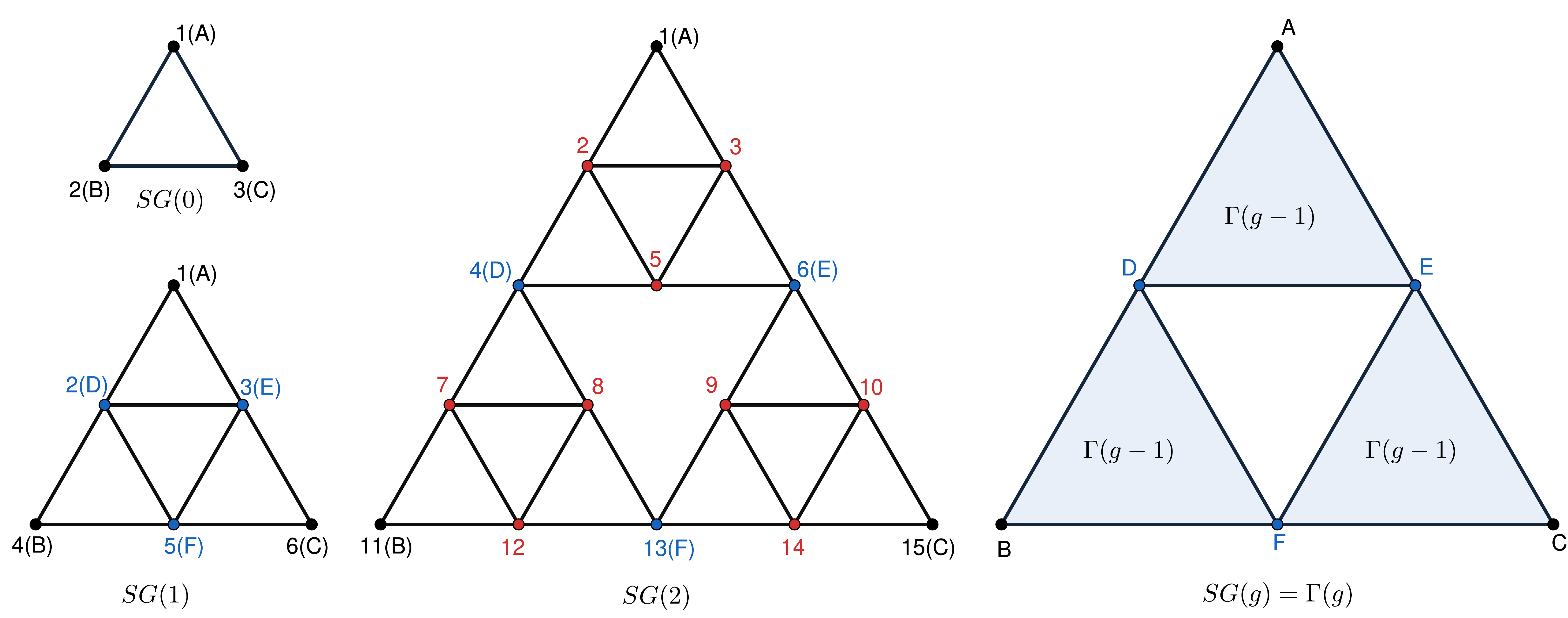}
\caption{Sierpinski Gasket: As shown in the figure, the initial network is equilateral triangle. The network $SG(1)$ and $SG(2)$ after two iterations are shown in the figure on the left, where the black nodes are the initial nodes, the blue nodes are the nodes generated after the first iteration, and the red nodes are the new nodes generated after the second iteration. The graph on the far right is the network after the $g$-th iteration, which can be assembled by three regions $\Gamma(g-1)$, namely $g-1$ generation network SG(g-1), reflecting the self-similarity of the network.}
\label{figl}
\end{figure}

After a brief introduction to the Sierpinski Gasket, we will define the multi-level horizontal partitioning method on the network $SG(g)$. First, we define the 1-level horizontal segmentation method of network $SG(g)$: the line passing through the nodes at the midpoints of line segment $AB$ and line segment $AC$ is taken as the $1$-level parting line, denoted as $l_1$, and the part below the parting line is retained while the rest is removed, in which the nodes on the spline are retained and the edges that coincide with the parting line are removed. Here, the rest of the network $SG(g)$ after being segmented by $1$-level parting line is denoted as $SSG(g,1)$. Next, the $2$-level parting line, denoted as $l_2$, is defined as a line that passes through two nodes, which are respectively located on the line segments $AB$ and $AC$, and the distance between them and node $A$ is $2^g-2^{g-2}$. Similarly, just keep the lower half of the divided network $SG(g)$ and denote it as $SSG(g,2)$. Finally, for any positive integer $k$, we can define the $k$-level horizontal parting line and its residual network after segmentation in a similar way: $k$-level horizontal parting line, denoted as $l_k$, is the straight line passing through two nodes, which are respectively located on the line segments $AB$ and $AC$, and the distance between them and node $A$ is $2^g-2^{g-k}$; the residual network after segmentation denoted as $SSG(g,k)$ is the lower part of k-level splitter line, where the nodes on the splitter line are retained, but the edges that coincide with the splitter line are not. Obviously, the segmentation parameter here must satisfy: $1\leq k\leq g$.

\begin{figure}[t]
\includegraphics[scale=0.25]{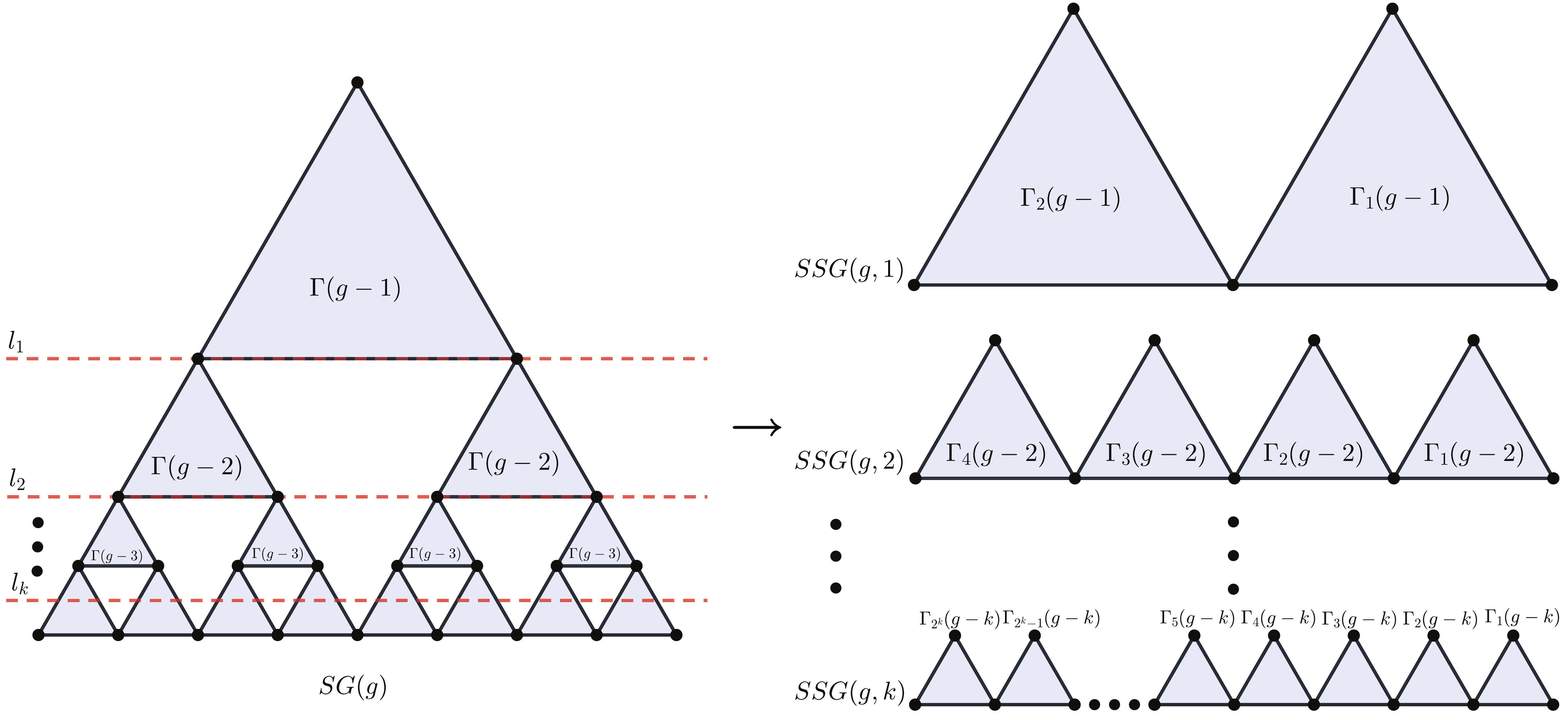}
\caption{Multi-level Horizontal Partitioning Method: in the figure above, the graph on the left is $g$ generation Sierpinski Gasket $SG(g)$, and the parting lines $l_1,l_2,\ldots,l_k$; the graphs on the right are the residual networks after segmentation according to the parting lines.}
\label{fig2}
\end{figure}

It can be found that network $SSG(g,k)$ is essentially connected by $2^k$  Sierpinski Gasket $SG(g-k)$. Therefore, the mean time to absorption(MTA) over the residual network $SSG(g,k)$ is based on the MTA over a sequence of connected Sierpinski Gaskets. We hereby define the auxiliary series network, denoted as $ASG(g,m)$, to concatenate $m$ identical Sierpinski Gasket $SG(g)$ in sequence as shown in Fig.3.

Since the auxiliary network $ASG(g,m)$ is composed of $m$ regions $\Gamma(g)$, we denote these $m$ regions as $\Gamma_1(g),\Gamma_2(g)\ldots,\Gamma_m(g)$ in order from right to left for clarity. And for any region, $\Gamma_j(g)(1\leq j\leq m)$, we represent the vertices at the three corners on the outside as $(A,j)$, $(B,j)$, and $(C,j)$ as shown in Fig.3. Similarly, to accurately represent the location of each node, we denote the node $i$ in region $\Gamma_j(g)$ as $(i,j)$. It is worth noting that using this notation results in certain nodes being tagged twice at the same time, in other words, different symbols representing the same node. For example, $(B,n)=(C,n+1)$ where $1\leq n<m$. However, this is not a defect, but a deliberate treatment for later calculations.

Then, we record the number of all nodes and the number of all edges in the auxiliary network $ASG(g,m)$ as $N_A(g,m)$ and $E_A(g,m)$, respectively. Based on the structure of the network, we can prove that:
$$
N_A(g,m)=\frac{m}{2}(3^{g+1}+1)+1\quad \textrm{and}\quad E_A(g,m)=m\cdot3^{g+1}.
$$
Moreover, the total number of nodes and the total number of edges on the residual network $SSG(g,k)$ are denoted as $N_S(g,k)$ and $E_S(g,k)$. The following equation can be obtained from the relationship between the residual network $SSG(g,k)$ and the auxiliary network $ASG(g,m)$:
$$
N_S(g,k)=2^{k-1}(3^{g-k+1}+1)+1\quad \textrm{and}\quad E_S(g,k)=2^k\cdot3^{g-k+1}.
$$

Here, we have explained the structure and some related properties of Sierpinski Gasket $SG(g)$, residual network after $k$-level horizontal partitioning $SSG(g,k)$, and auxiliary network $ASG(g,m)$. Next we will discuss the mean time to absorption on the network.

\begin{figure*}[t]
\centering
\includegraphics[scale=0.25]{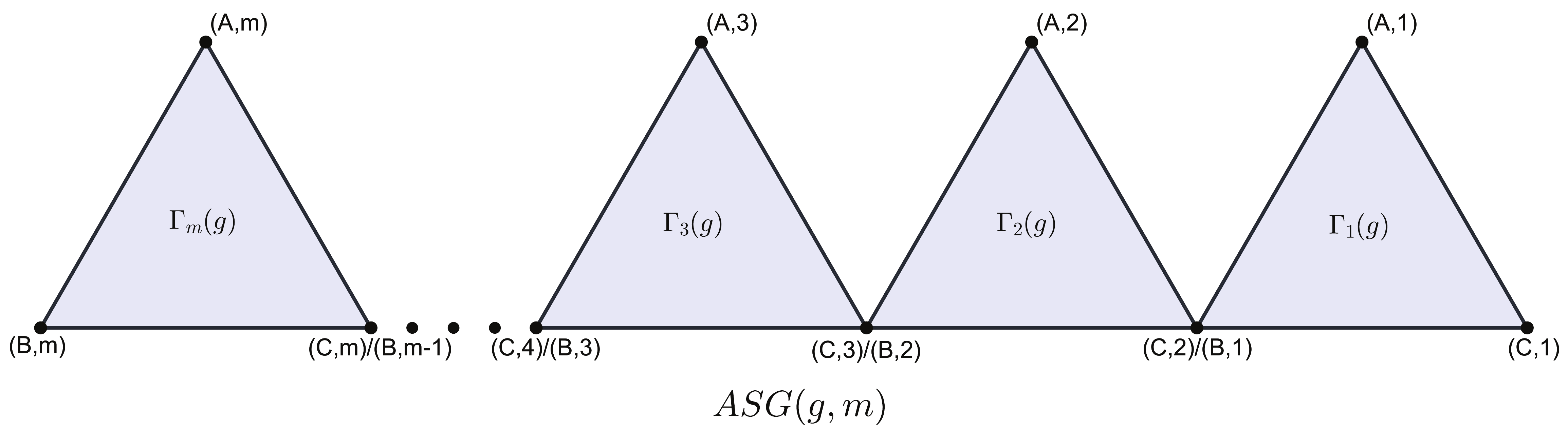}
\caption{Auxiliary Network $ASG(g,m)$: This network can be obtained by connecting $m$ Sierpinski Gaskets $SG(g)$ in sequence.}
\label{fig3}
\end{figure*}

\section{Analytical Expression of Mean Time to Absorption on network $SSG(g,m)$}

In this section, we calculate the mean time to absorption over the residual network after $k$-level horizontal partitioning $SSG(g,k)$. In order to facilitate the presentation of the calculation process, this section will be divided into the following 4 subsections.

\subsection{Mean Time to Absorption on Network}

In this subsection, the random walk process on the network and the definition of the mean time to absorption (MTA) will be introduced respectively as preliminary knowledge of subsequent calculation process.

First, the unbiased Markovian random walk on the network $G$ will be presented as the basis for the definition of the mean time to absorption, where $G$ generally refers to a general undirected and unweighted connected network.
The total number of nodes and the total number of edges in the network $G$ are denoted as $N$ and $E$ respectively, and each node is numbered from $1$ to $N$. Without loss of generality, we set node $\alpha$ as the absorptive node. Starting from any site $i$ other than the absorptive node $\alpha$, the walker can jump to any of its nearest-neighbor nodes with equal probabilities at each time step(taken by unity). But once the walker enters the absorptive node $\alpha$, it will stop random walk and no longer move. Therefore, the probability of transition at each step is the reciprocal of the degree of the node without absorptive node $\alpha$. Here, the degree of site $i$ can be denoted as $d_i$, then, the probability can be expressed as follow:
\begin{eqnarray*}
p_{ij}=
\left\{
\begin{array}{lr}
\frac{1}{d_i},\quad \textrm{if $i\sim j$ and $i\neq \alpha$}\\
0,\quad \textrm{others}
\end{array}
\right.
\end{eqnarray*}
where $p_{ij}$ is the probability that the walker jump from the site $i$ to the site $j$, and $i\sim j$ means that the site $i$ is directly connected to the site $j$.
Furthermore, when no absorptive node is set in network $G$, every node in the network can be traversed by the random walk Markov chain, that is, for any initial node, walker will reach any node in the network with probability 1. Therefore, after setting the absorptive node again, walker will be captured by the absorptive node at some point. As the size of the network approaches infinity, namely $N\rightarrow+\infty$, the above conclusion is still true, even if the mean time to absorption(MTA) approaches infinity.

Then, the mean first-passage time(MFPT), denoted as $T_{i,j}$, is defined as the mean time of the walker starting from node i and reaching the target node for the first time through a random walk.
Here, when the starting point and the ending point are the same, it is stipulated that: $T_{i,i}=0$.
 The average time to absorption of node $i$ on the network $G$ with absorptive node $\alpha$, denoted as $T_i$, is defined as the average time for walker to start from node $i$ and finally enter the absorptive node and stop random walker.
In fact, it is easy to prove by the above definition: $T_{i,\alpha}=T_i$ if $\alpha$ is the only absorptive node in the network.
Then the total time to absorption on the network, denoted as $T_{total}$, is defined as:
$$
T_{total}=\sum_{i\in\Omega}T_{i}=\sum_{i\in\Omega}T_{i,\alpha}
$$
where, $\Omega$ be used to represented the set of all nodes in network. Therefore, the mean time to absorption (MTA) of network $G$, denoted as $\bar T$, is defined as:
$$
\bar T=\frac{T_{total}}{N-1},
$$
where $N$ refers to the total number of node. Moreover, the MTA on a network with multiple absorptive nodes can naturally be generalized by the above definition, and it does not need to be explained in detail here.

In this paper, we set the corner node $C$ of the Sierpinski Gasket $SG(g)$ as the absorptive node. According to the results of previous published articles, it can be seen that the analytical expressions of the total time to absorption and the mean time to absorption on the Sierpinski Gasket $SG(g)$ are determined by the iterative variable $g$, and respectively meet the following relationships\cite{meyer2012exact}:
$$
T_{total}(g)=\frac52\cdot3^g\cdot5^g+2\cdot5^g-\frac12\cdot3^g
$$
and
$$
\bar T(g)=\frac{1}{N(g)-1}T_{total}(g)=\frac{3^g\cdot5^{g+1}+4\cdot5^g-3^g}{3^{g+1}+1}.
$$
Then, our goal is to calculate the analytical expression of the mean time to absorption on the residual network after $k$-level horizontal partitioning $SSG(g,k)$ with absorptive node $C$.

It is worth mentioning that for any network $G$, if its corresponding transition probability matrix is known, then the MTA on the network can be calculated by matrix algorithm \cite{zhang2020mean}.
Naturally, the above method can be used to calculate the MTA in the Sierpinski Gasket $SG(g)$, residual network after $k$-level horizontal partitioning $SSG(g,k)$, and auxiliary network $ASG(g,m)$. However, since the rank of the matrix corresponding to the above three networks is $N(g)-1$, $N_S(g)-1$ and $N_A(g)-1$ respectively, when the number of iterations $g$ approaches infinity, the rank of the matrix will also approach infinity. Therefore, although the above matrix algorithm can be used to solve the problem of MTA of any node in a simple connected network of any finite scale, it cannot deal with the situation when the network scale is very large. It is necessary to present another way to calculate the mean time to absorption and make it easier to analyze in the residual network after $k$-level horizontal partitioning $SSG(g,k)$

\subsection{MTA of $ASG(g,m)$}

As mentioned earlier, although the MTA on the Sierpinski gasket network $SG(g)$ has been discussed, the conclusions are not enough to support the following calculations.\cite{meyer2012exact} Therefore, in this section, we will firstly present the analytical expressions of MTA on Sierpinski gasket network with two and three outermost absorptive nodes based on these results. These expressions will be the basic prerequisites for subsequent calculations.

First, $\Omega(g)$ is defined as the set of all nodes in the network $SG(g)$, and the three nodes in the outermost corner are marked as $A$, $B$ and $C$, as shown in Fig.1.
If $B$ and $C$ nodes are set as absorptive nodes on network $SG(g)$, then the mean first passage time from node $i$ to the absorptive nodes, the total time to absorptions and the mean time to absorptions are denoted as $T^2_i(g)$, $T^2_{total}(g)$ and $\bar T^2(g)$, respectively.
Similarly, if the three nodes $A$, $B$ and $C$ are set as absorptive nodes, then the mean first passage time from node $i$ to the absorptive nodes, the total time to absorptions and the mean time to absorptions are denoted as $T^3_i(g)$, $T^3_{total}(g)$ and $\bar T^3(g)$, respectively.

According to the symmetry of the Sierpinski Gasket network, the following relationship can be obtained:
\begin{align}
T^2_A(g)&=5T'(g),\label{eq:it}\\
T^2_{total}(g)&=3^g\cdot5^g+\frac12(5^g-3^g).\label{eq:T2t}
\end{align}
Then, the mean time to absorptions on $SG(g)$ with double absorptive nodes satisfies:
\begin{eqnarray*}
\bar T^2(g)=\frac{1}{N(g)-2}T^2_{total}(g)=\frac{2\cdot3^g\cdot5^g+5^g-3^g}{3^{g+1}-1}.
\end{eqnarray*}
{Due to the limited space, the specific derivation process of the above conclusions is presented in the Appendix.}

After solving the problem of mean time to absorptions with multiple absorptive nodes on Sierpinski Gasket $SG(g)$, we then calculate the MTA on auxiliary network $ASG(g,m)$ with absorption nodes $(C,1)$. Given that network $ASG(g,m)$ is connected by $m$ regions $\Gamma(g)$, the set of all nodes on network $ASG(g,m)$ is denoted as $\Omega_A$, the set of all nodes on region $\Gamma_j(g)(1\leq j\leq m)$ is denoted as $\Omega_A^j$, and the set only containing nodes $(B,j)$ and $(C,j)$ is denoted as $\hat\Omega_A^j$. Furthermore, let $\hat\Omega_A=\hat\Omega_A^1\cup\hat\Omega_A^2\cup\ldots\cup\hat\Omega_A^m$, $\bar\Omega_A^j=\Omega_A^j/\hat\Omega_A^j$ and $\bar\Omega_A=\Omega_A/\hat\Omega_A$. The MFTP from node $(a,b)$ to node $(c,d)$ in the auxiliary network $ASG(g,m)$ is denoted as $T^A_{(a,b),(c,d)}(g,m)$. Then, the total time to absorption on the auxiliary network $ASG(g,m)$ is denoted as $T^A_{total}(g,m)$, and the MTA is denoted as $\bar T^A(g,m)$.

Based on the above node classification, we can make the following analysis: For any node $(i,j)\in\bar\Omega_A$, the path to the absorptive node $(C,1)$ can be divided into two segments. First of all, starting from node $(i,j)$, walker reach nodes $(B,j)$ or $(C,j)$ through mean time $T^2_i(g)$. Then walker starts at $(B,j)$ or $(C,j)$, and finally arrives at the absorptive node $(C,1)$. Therefore, $T^A_{total}(g,m)$ satisfies the following relation:
\begin{eqnarray}\label{eq:TAt0}
T^A_{total}(g,m)&=&\sum_{(i,j)\in\Omega_A}T^A_{(i,j),(C,1)}(g,m)\nonumber\\
&=&\sum_{j=1}^{m}\sum_{(i,j)\in\bar\Omega^j_A}T^A_{(i,j),(C,1)}(g,m)+\sum_{(i,j)\in\hat\Omega_A}T^A_{(i,j),(C,1)}(g,m).
\end{eqnarray}
If the start node $(i,j)$ is randomly selected from set $\bar\Omega_A^j$ according to the uniform distribution, based on the symmetry of nodes $(B,j)$ and $(C,j)$ in region $\Gamma_j(g)$, the walker starts from the start node and arrives at any node in $(B,j)$ or $(C,j)$ with equal probability. Therefore, for any $j\in[1,m]$, it can be deduced that:
\begin{eqnarray*}
&&\sum_{(i,j)\in\bar\Omega^j_A}T^A_{(i,j),(C,1)}(g,m)\\
&=&\sum_{(i,j)\in\bar\Omega^j_A}T^2_{(i,j)}(g)+\sum_{(i,j)\in\bar\Omega^j_A}\big[\frac12T^A_{(B,j),(C,1)}(g,m)+\frac12T^A_{(C,j),(C,1)}(g,m)\big]\\
&=&T^2_{total}(g)+\frac12(N(g)-2)\big[T^A_{(B,j),(C,1)}(g,m)+T^A_{(C,j),(C,1)}(g,m)\big]
\end{eqnarray*}
Then, substitute the above equation into Eq.(\ref{eq:TAt0}), and it can be obtained that:
\begin{eqnarray*}
T^A_{total}(g,m)&=&m\cdot T^2_{total}(g)+\frac{N(g)-2}{2}\sum_{j=1}^m\big[T^A_{(B,j),(C,1)}(g,m)\\
&&+T^A_{(C,j),(C,1)}(g,m)\big]+\sum_{(i,j)\in\hat\Omega_A}T^A_{(i,j),(C,1)}(g,m).
\end{eqnarray*}
As mentioned above, node $(B,j)$ and node $(C,j-1)$ both represent the same node where $1<j\leq m$, so the above formula can be simplified as:
\begin{eqnarray}\label{eq:TAt1}
T^A_{total}(g,m)&=&m\cdot T^2_{total}(g)+(N(g)-2)\Big[\sum_{(i,j)\in\hat\Omega_A}T^A_{(i,j),(C,1)}(g,m)\nonumber\\
&&-\frac12T^A_{(B,m),(C,1)}(g,m)\Big]+\sum_{(i,j)\in\hat\Omega_A}T^A_{(i,j),(C,1)}(g,m)\nonumber\\
&=&m\cdot T^2_{total}(g)+(N(g)-1)\sum_{(i,j)\in\hat\Omega_A}T^A_{(i,j),(C,1)}(g,m)\nonumber\\
&&-\frac{N(g)-2}{2}T^A_{(B,m),(C,1)}(g,m).
\end{eqnarray}
Naturally, we have to figure out $T^A_{(i,j),(C,1)}(g,m)$ for any node $(i,j)\in\hat\Omega_A$. Since $T^A_{(C,1),(C,1)}(g,m)=0$, we only need to consider nodes $(B,j)$ where $j\in[1,m]$. According to the property of random walk, starting from node $(B,x)$, where $1<x<m$, the walker will appear at node $(B,x-1)$ or node $(B,x+1)$ for the first time with equal probability, and the average time of this process will be $T_{A,C}(g)$. Therefore, when only nodes $(i,j)\in\hat\Omega_A$ is considered, we can reduce this model to a random walk model on a one-dimensional finite lattice with a absorptive node, where the time of each random walk is $T_{A,C}(g)$.

Since the time of each random walk is only equably varied, we only need to establish a one-dimensional lattice denoted as $L(m)$ with a length of $m$, in which the length of each edge is the unit. Thus, for $m+1$ nodes on the lattice, we numbered them sequentially from $0$ to $m$, and set node $0$ as the absorptive node. Then, we use $T_{a\leftrightarrow b}(m)$ to represent the expected commute time on lattice $L(m)$. According to the effective resistance principle, it can be obtained that\cite{chandra1996electrical}:
$$
T_{a\leftrightarrow b}(m)=T^L_{a,b}(m)+T^L_{b,a}(m)=2mR_{a,b},
$$
where $T^L_{a,b}(m)$ represent the MFPT from node $a$ to node $b$ on lattice $L(m)$ and $R_{a,b}$ the effective resistance between node $a$ and node $b$ on the equivalent resistance network $L(m)$. According to the properties of one-dimensional lattice network, it is easy to know: $R_{a,b}=b-a$. Setting $a=0$, we can obtain that:
$$
T_{0\leftrightarrow b}(m)=T^L_{0,b}(m)+T^L_{b,0}(m)=2mR_{0,b}=2mb.
$$
In addition, it can be obtained from the network structure: $T^L_{0,b}(m)=T^L_{0,b}(b)=T^L_{b,0}(b)$. Therefore, the following relationship holds:
$$
T^L_{0,b}(m)=T^L_{0,b}(b)=\frac12 T_{0\leftrightarrow b}(m)=b^2.
$$
On this basis, it can be solved that:
$$
T^L_{b,0}(m)=2mb-T^L_{0,b}(m)=2mb-b^2.
$$
Based on the above conclusions and the previous analysis, it can be proofed that:
$$
T^A_{(B,m),(C,1)}(g,m)=T^L_{m,0}(m)\cdot T_{A,C}(g)=2\cdot m^2\cdot 5^g
$$
and
\begin{eqnarray*}\label{eq:TAtotal}
\sum_{(i,j)\in\hat\Omega_A}T^A_{(i,j),(C,1)}(g,m)&=&\sum_{b=1}^m T^L_{b,0}(m)\cdot T_{A,C}(g)=2\cdot5^g\sum_{b=1}^m(2mb-b^2)\\
&=&\frac13m(m+1)(4m-1)5^g.
\end{eqnarray*}
Substitute Eq.(\ref{eq:T2t}) and the above expression into Eq.(\ref{eq:TAt1}) to obtain:
\begin{eqnarray}
T^A_{total}(g,m)=(2m^3+\frac12m)\cdot3^g\cdot5^g
+(\frac23m^3+m^2+\frac13m)\cdot5^g-\frac{m}{2}\cdot3^g.
\end{eqnarray}
Since the total time to absorption in the auxiliary network has been calculated, the MTA can be derived as follows:
\begin{eqnarray}
\bar T^A(g,m)&=&\frac{T^A_{total}(g,m)}{N_A(g,m)-1}\nonumber\\
&=&\frac{1}{3m(3^{g+1}+1)}\Big((12m^3+3m)\cdot3^g\cdot5^g+(4m^3+6m^2+2m)\cdot5^g-3m\cdot3^g\Big).
\end{eqnarray}

\subsection{MTA of $SSG(g,k)$}
Based on the total time to absorption on the auxiliary network $ASG(g,m)$, we can obtain the expression of the total time to absorption on the residual network after $k$-level horizontal partitioning $SSG(g,k)$, denoted as $T^S_{total}(g,m)$, as follows:
\begin{eqnarray*}
T^S_{total}(g,k)&=&T^A_{total}(g-k,2^k)\\
&=&(2^{3k+1}+2^{k-1})\cdot15^{g-k}+(\frac132^{3k+1}+2^{2k}+\frac132^k)\cdot5^{g-k}-2^{k-1}\cdot3^{g-k}.
\end{eqnarray*}
Then, it can be obtained that the MTA of $SSG(g,k)$, denoted as $\bar T^S(g,k)$, satisfies the following equation:
\begin{eqnarray}\label{eq:TS}
\bar T^S(g,k)&=&\frac{T^S_{total}(g,k)}{N_S(g,k)-1}\nonumber\\
&=&(4^{k+1}+1)\frac{15^{g-k}}{3^{g-k+1}+1}+(\frac134^{k+1}+2^{k+1}+\frac23)\frac{5^{g-k}}{3^{g-k+1}+1}-\frac{3^{g-k}}{3^{g-k+1}+1}.
\end{eqnarray}
Therefore, we finally figured out the analytical expression of the mean time to absorption on the residual network after $k$-level horizontal partitioning $SSG(g,k)$ when node $C$ is the absorptive node. Letting $k=0$, we can verify that $\bar T^S(g,0)=\bar T(g)$. It is worth noting that, by observing the above expression, we can find that the leading term of $\bar T^S(g,k)$ obeys:
\begin{eqnarray}\label{eq:TS-a}
\bar T^S(g,k)\sim(4^{k+1}+1)\frac{15^{g-k}}{3^{g-k+1}+1}\sim\frac{4}{3}\cdot\big(\frac{4}{5}\big)^k\cdot5^g.
\end{eqnarray}
Therefore, the mean time $\bar T^S(g,k)$ is a monotonically increasing function of the number of iterations $g$, and a monotonically decreasing function of the partition coefficient $k$.

\begin{figure}[t]
\centering
\begin{minipage}{0.32\linewidth}
\centering
\includegraphics[scale=0.28]{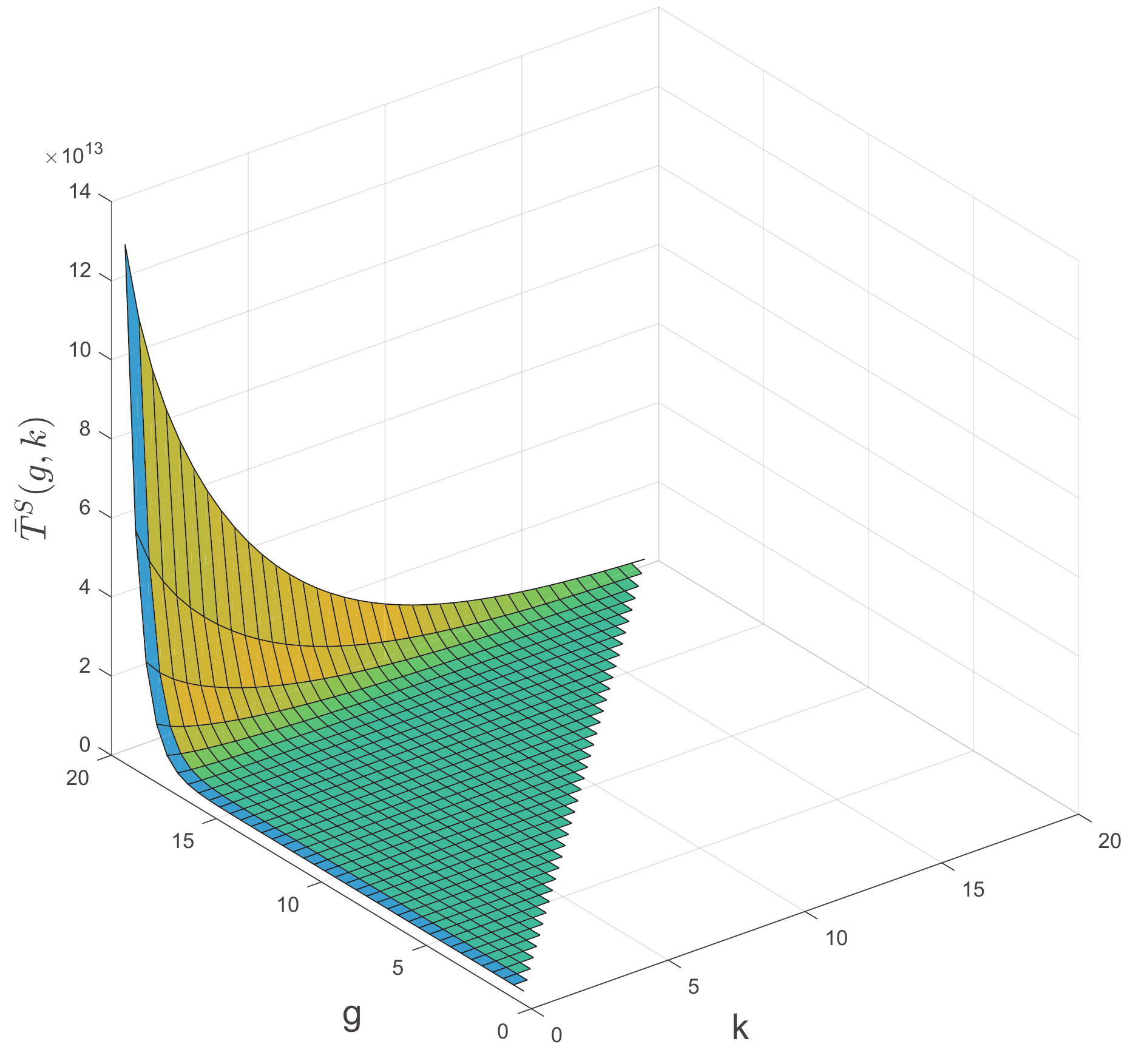}
\caption{Numerical simulation diagram of $\bar T^S(g,m)$.}
\end{minipage}\label{fig4}
\centering
\begin{minipage}{0.32\linewidth}
\centering
\includegraphics[scale=0.28]{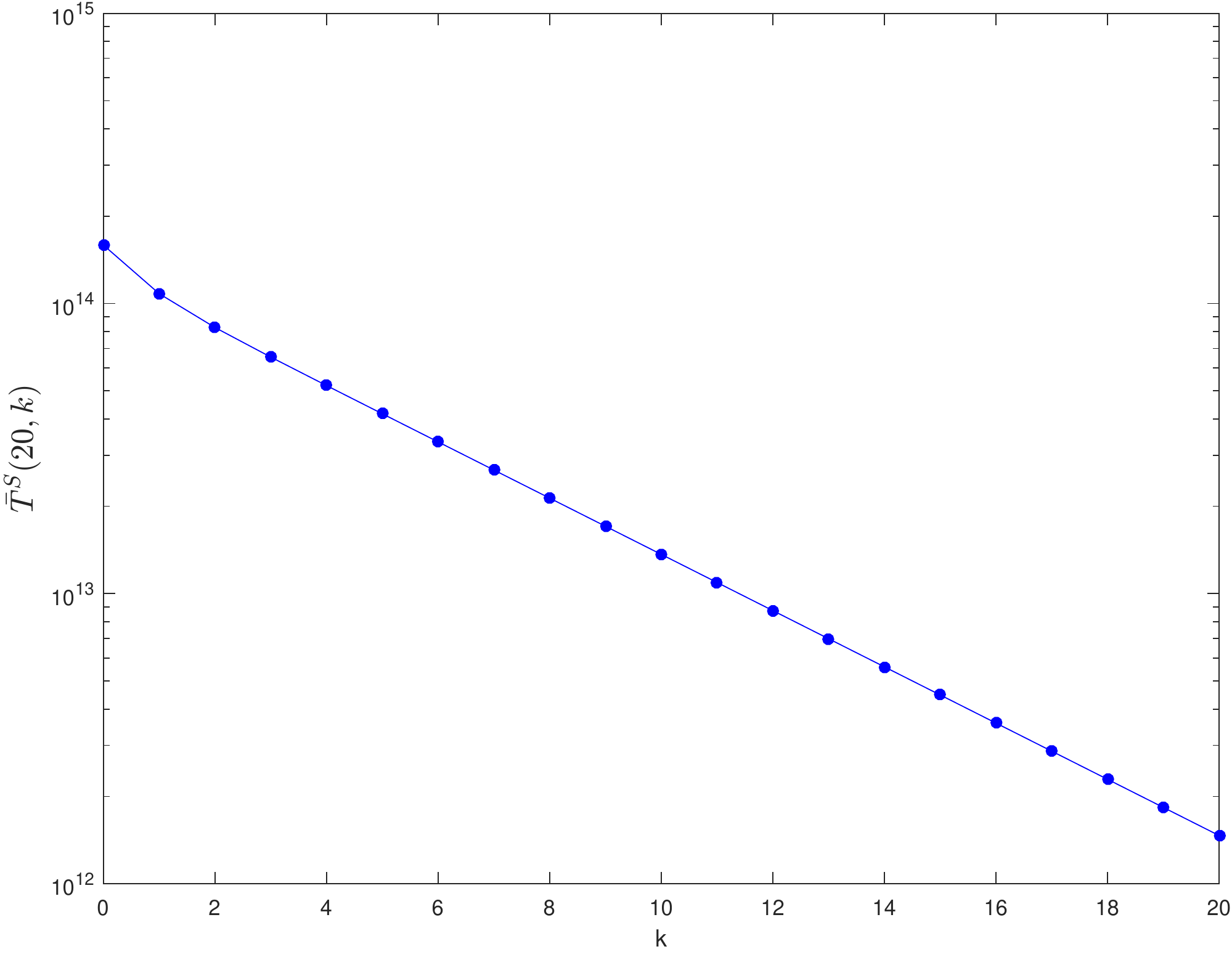}
\caption{When the number of iterations $g=20$, the scatter diagram of mean time to absorption $\bar T^S(20,k)$ with respect to the change of segmentation parameter $k$.}
\end{minipage}\label{fig5}
\centering
\begin{minipage}{0.32\linewidth}
\centering
\includegraphics[scale=0.28]{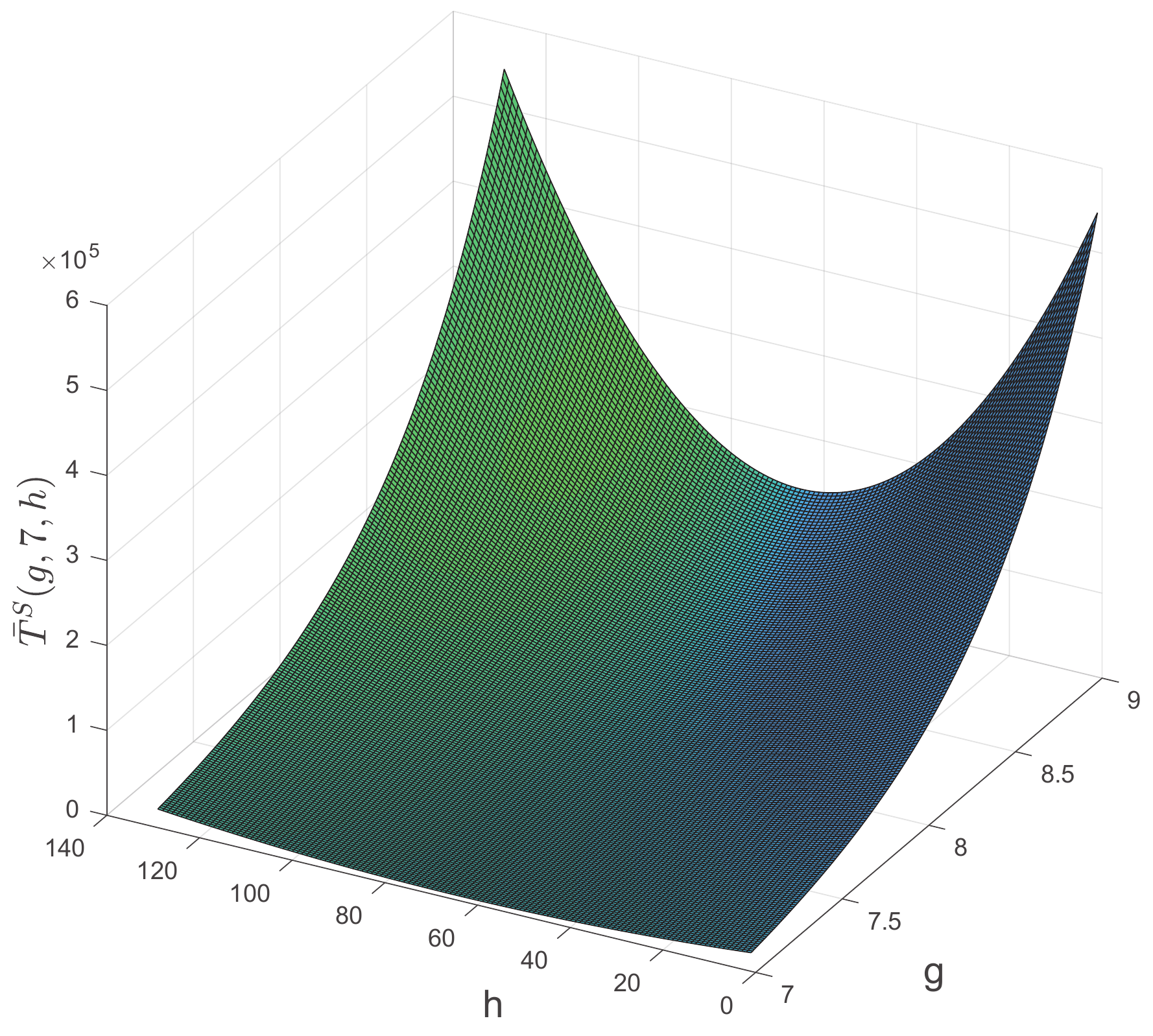}
\caption{When the segmentation parameter $k=7$, the scatter diagram of mean time to absorption $\bar T^S(g,7,h)$ with respect to the change of position coefficient $h$ and number of iterations $g$.}
\end{minipage}\label{fig6}
\end{figure}

In order to more intuitively show the changing trend of $\bar T^S(g,k)$ as the number of iterations $g$ and the partition coefficient $k$ increase, we performed a numerical simulation on $\bar T^S(g,k)$ by Matlab, and the results are shown in Fig.\ref{fig4}. Obviously, as the number of iterations increases, the MTA on the network $SSG(g,k)$ increases exponentially, and as the partition coefficient $k$ increases, although the MTA will decrease exponentially. Therefore, the partition coefficient $k$ has an obvious negative correlation with the MTA on the network $SSG(g,k)$.
In order to better demonstrate this process, we draw the Fig.\ref{fig5}, where we fixed the number of iterations $g=20$. It can be seen that with the increase of segmentation parameter $k$, the mean time $\bar T^S(g,k)$ decreases exponentially, which is also consistent with previous analysis. When $k=g$, $\bar T^S(g,k)$ is the minimum. Therefore, it is easy to prove that $\bar T^S(g,g)$ satisfies the following relation:
\begin{eqnarray}
\bar T^S(g,g)=\frac13\cdot4^{g+1}+2^{g-1}+\frac16\sim \big(N_s(g,g)\big)^2\sim \big(\bar T(g)\big)^{\frac{\ln4}{\ln5}}.
\end{eqnarray}
Moreover, when $k=0$, $\bar T^S(g,k)$ has a maximum value and satisfies:
\begin{eqnarray}
\bar T^S(0,g)=\frac{3^g\cdot5^{g+1}+4\cdot5^g-3^g}{3^{g+1}+1}=\bar T(g)
\sim \big(N_s(g,0)\big)^{\frac{\ln5}{\ln3}}.
\end{eqnarray}

\subsection{MTA of $SSG(g,m)$ with other absorptive node}

In this section, we will analyze the analytical expression of the MTA with respect to the position of the absorptive node. In order to parameterize the position of the absorptive node, we make the following provisions: the node $(B, m)$ in the auxiliary network $ASG(g, m)$ is also marked as $(C, m+1)$, and when the node $(C, h)$ is set as the absorptive node, the position coefficient is recorded as $h(1\le h\le m+1)$. In the same way, node $B$ in the horizontal partitioned Sierpinski Gasket network $SSG(g,k)$ is denoted as $(C, 2^k+1)$, other nodes use the same label as in the auxiliary network, and the position coefficient of absorptive node $(C, h)$ is denoted as $h(1\le h\le 2^k+1)$.

First of all, the case of $(C, h)$ as an absorptive node in the auxiliary network $ASG(g, m)$ is considered. We split the auxiliary network at the absorptive node $(C, h)$. Obviously at this time, the left side of node $(C, h)$ is an auxiliary network $ASG(g, m-h+1)$ composed of $m-h+1$ self-similar regions, and the right side is an auxiliary network $ASG(g, h-1)$ composed of $h-1$ regions, and the left and right regions have no common nodes except for the absorptive node $(C, h)$. The total absorptive time on the network $ASG(g, m)$ with the absorptive node $(C, h)$ is denoted as $T^A_{total}(g,m,h)$, then from equation (\ref{eq:TAtotal}) we can obtain:
\begin{eqnarray}
T^A_{total}(g,m,h)
&=&T^A_{total}(g, m-h+1)+T^A_{total}(g,h-1)\nonumber\\
&=&\big[2m^3-6m^2h+6mh^2+6m^2-12mh+\frac{13}{2}m\big]\cdot 3^g\cdot 5^g\nonumber\\
&&+\big[\frac{2}{3}m^3-2m^2h+2mh^2+3m^2-6mh+\nonumber\\
&&\frac{16}{3}m+2h^2+2-4h\big]\cdot 5^g-\frac{1}{2}m\cdot 3^g
\end{eqnarray}

Based on the total absorptive time on the auxiliary network, the total absorptive time  on the network $SSG(g,k)$ with the absorptive node $(C, h)$, denoted as $T^S_{total}(g,k,h)~(1\le h\le 2^k+1)$, can naturally be deduced:
\begin{eqnarray}
T^S_{total}(g,k,h)&=&T^A_{total}(g-k, 2^k, h)\nonumber\\
&=&\big[ 2^{3 k+1}-6 \cdot 2^{2 k} h+6 \cdot 2^{k} \cdot h^{2}+6\cdot2^{2k}-12 \cdot 2^{k} \cdot h+\frac{13}{2} \cdot 2^{k}\big]\nonumber\\
&&\cdot 3^{g-k}\cdot 5^{g-k}+\big[\frac{2}{3} \cdot 2^{3 k}-2^{2k+1} \cdot h+2^{k+1} h^{2}+3\cdot2^{2 k}\nonumber\\
&&-6 \cdot 2^{k} \cdot h+2 h^{2}+2-4 h\big]\cdot 5^{g-k}-\frac{1}{2}\cdot 2^k\cdot3^{g-k}
\end{eqnarray}

Then, the analytical expression of the MTA with the absorptive node position coefficient $h$ in the network $SSG(g,k)$, denoted as $\bar{T}^S(g,k,h)$, satisfies:
\begin{eqnarray}\label{eq:MTS}
\bar T^S(g,k,h)&=&\frac{T^S_{total}(g,k,h)}{N_S(g,k)-1}\nonumber\\
&=&\big[12h^2-(24+6\cdot 2^{k+1})h+2^{2k+2}+6\cdot 2^{k+1}+13\big]\frac{5^{g-k}\cdot 3^{g-k}}{3^{g-k+1}+1}\nonumber\\
&&+\big[(4+2^{2-k})h^2-(2^{k+2}+12+2^{3-k})h+\frac{2^{2k+2}}{3}+3\cdot2^{k+1}\nonumber\\
&&+2^{2-k}\big]\frac{5^{g-k}}{3^{g-k+1}+1}+\frac{3^{g-k}}{3^{g-k+1}+1}
\end{eqnarray}
In order to estimate the position of the absorptive node that makes the MTA take the minimum, we let:
$$
f(h)=12h^2-(24+6\cdot 2^{k+1})h+2^{2k+2}+6\cdot 2^{k+1}+13
$$
By deriving $f(h)$, we know that when $h=2^{k-1}+1$, the main term of $\bar T^S(g,k,h)$ will take the minimum value. Therefore, it can be estimated that when the number of iterations $g$ is large enough and the position of the absorptive node is set to $(C,2^{k-1} +1)$, the MTA is the smallest, that is, the absorption efficiency of the node $(C,2^{k-1}+1)$ is the highest in nodes $\{(C,1), (C,2),\ldots,(C,2^k+1)\}$.
Eq.(\ref{eq:MTS}) has been numerically simulated, and the results are shown in Fig.\ref{fig6}. Obviously, when the position coefficient of the absorptive node is $2^{k-1}+1$, the MTA is the smallest, which is consistent with our estimation.

\section{Conclusion}

In summary, the analytical expressions and approximate expressions of MTA on the Horizontal Partitioned Sierpinski Gasket network model are obtained in this article. It can be seen from the segmentation structure of the network that the partition coefficient $k$ not only indicates the degree of incompleteness of the network $SSG(g,k)$, but also an indicator of the size of the self-similar structure in the remaining network. And the larger the $k$ is, the smaller the self-similar structure of the remaining network $SSG(g,k)$ is. Then the approximate expression of MTA in Eq.(\ref{eq:TS-a}) fully shows that when the partition coefficient $k$ increases, $\bar T^S(g,k)$ will decrease exponentially. However, when the partition coefficient $k$ satisfies $k\ll g$, the MTAs of the original network and the segmented network will have the same exponential growth rate with the number of iterations $g$. Therefore, we can conclude that in a network with a local self-similar structure, the size of these self-similar structures directly restricts the influence of self-similarity on the properties of random walks. Finally, we found the position of the absorptive node with the highest absorption efficiency by parameterizing the position of the node.

\section*{Acknowledgment}

This work was supported by National Natural Science Foundations of China Grant (No.12026214, No.11871061 and No.12026213), Natural Science Research Major Project of Higher Education in Jiangsu Province (No. 17KJA120002) and the 333 Project of Jangsu Province.

\section*{DATA AVAILABILITY}

The data that support the findings of this study are available from the corresponding author upon reasonable request.

\appendix
\section{MTA with multiple absorptive nodes}

Firstly, we consider the iterative relationship of random walks of nodes corresponding to the first-generation network in the network $SG(g)$.
Following the mark in Fig.1, the mean first passage time of node $A$ to nodes $D$ or $E$ is denoted as $T'(g)$, that is, the mean first passage time of node $A$ to absorptions when $D$ and $E$ are set as the absorptive nodes in the network $SG(g)$. Then, when $B$ and $C$ in network $SG(g)$ are set as absorptive nodes, the average time to absorptions of nodes in $\Omega(1)$ are: $T^2_A(g)$, $T^2_D(g)$, $T^2_E(g)$ and $T^2_F(g)$.
Based on the unbiased Markovian random walk definition and the symmetry on the gasket network, the following equations between the mean value of the first passage time can be established:
\begin{eqnarray*}
\left\{
\begin{array}{lr}
T^2_A(g)=\frac12(T'(g)+T^2_D(g))+\frac12(T'(g)+T^2_{E}(g))\\
T^2_D(g)=\frac14(T'(g)+T^2_A(g))+\frac14(T'(g)+T^2_E(g))\\
~~~~~~~~~~~+\frac14(T'(g)+T^2_F(g))+\frac14T'(g)\\
T^2_F(g)=\frac14(T'(g)+T^2_D(g))+\frac14(T'(g)+F_E(g))+\frac12T'(g)\\
T^2_E(g)=T^2_D(g)
\end{array}
\right.
\end{eqnarray*}

By solving the above equations, it is easy to prove that the following relations are true:
\begin{eqnarray}
T^2_A(g)=5T'(g).
\end{eqnarray}
According to the self-similarity of the network $SG(g)$, the area between the three points $A$, $D$ and $E$ in the network $SG(t)$ is $\Gamma(g-1)$, which corresponds to the network $SG(g-1)$ one by one. Therefore, it can be obtained that $T'(g)$ is also the first passage time of particles from node $A$ to node $B$ or $C$ in the $g-1$ generation, namely $T'(g)=T^2_A(g-1)$.
In the initial network $SG(0)$, it is easy to prove that the average time from node $A$ to nodes $B$ or $C$ for the first time is $1$, that is $T^2_A(0)=1$.
Then, from the iterative relationship Eq.(\ref{eq:it}) and initial conditions, it can be obtained that when nodes $B$ and $C$ in the network $SG(g)$ are set as absorptive nodes, the analytical expression of the mean time to absorption of node $A$ satisfies:
$$T^2_A(g)=5^g.$$
Moreover, according to the symmetry of the Sierpinski Gasket network and the definition of random walk on the network, it is known that the walker starting from node $A$ will reach node $B$ or $C$ for the first time after an average time $T^2_A(g)$ where the probability of receiving the walker at node $B$ and $C$ is the same.
Therefore, the mean first passage time for the walker to node $C$ from node $A$ must satisfy the following relation:
$$
T_{A,C}(g)=\frac12T^2_A(g)+\frac12(T^2_A(g)+T_{B,C}(g)).
$$
From the symmetry of the network $SG(g)$, we can also prove that: $T_{A,C}(g)=T_{B,C}(g)$. Therefore, it can be solved as:
$$
T_{A,C}(g)=T_{B,C}(g)=2\cdot T^2_A(g)=2\cdot5^g.
$$

Then, the relationship between $T^3_{total}(g)$ and $T_{total}(g)$ will be explained from the perspective of the random walk process.
The path of walker starting from any node $i\in\Omega(g)/\{A,B,C\}$ in the network $SG(g)$ to absorptive node $C$ can be divided into the following two types:

(1) The outermost nodes $A$ and $B$ do not exist in this path;

(2) The outermost node $A$ or $B$ exists in this path. Since node $A$ or $C$ may appear in the path multiple times, we only consider the outermost node that appears for the first time and divide the path into two segments at this node, where the walkers will go through the first path to reach the outermost $A$ or $C$ for the first time, and then start from the outermost node to reach the absorptive node $C$ through the second path.

Moreover, when the three outermost nodes $A$, $B$, $C$ in the network are all set as absorptive nodes and a node is randomly selected from $\Omega(g)/\{A,B,C\}$ according to the uniform probability distribution as the initial node, the walker starting from this node will arrive at any one of the three outermost nodes with equal probability after a random walk, due to the symmetry of the network.
Therefore, any path from node $i$ to outermost node $C$ can be divided into two parts where the first part can be regarded as the path of the node $i$ to the absorptive node in the network where the three outermost nodes are all set as absorptive nodes, and the second part is regarded as the path from the outermost node to node $C$. (If the end node of the first part of the path is $C$, the length of the second path is $0$.)
Based on the above analysis, we can conclude that the following relationship holds:
\begin{eqnarray*}
T_{total}(g)&=&\sum_{i\in\Omega(g)}T_{i,C}(g)\\
&=&\sum_{i\in\Omega(g)}\big[T^3_i(g)+(1-\delta_{n,C})T_{A,C}(g)\big]\\
&=&\sum_{i\in\Omega(g)}T^3_i(g)+2\big[\frac13(N(g)-3)+1\big]T_{A,C}(g)\\
&=&T^3_{total}(g)+\frac43N(t)T^2_A(g),
\end{eqnarray*}
where, $n\in\{A,B,C\}$ is the node of the absorbing walker starting from node $i$ when $\{A,B,C\}$ are set as the absorptive nodes and $\delta_{n,C}=1$ if $n=C$, otherwise $\delta_{n,C}=0$. Since $T_{total}(g)$ is known, it can be obtained that $T^3_{total}(g)$ and $\bar T^3(g)$ satisfy:
\begin{eqnarray*}
T^3_{total}(g)&=&T_{total}(g)-\frac{4}{3}N(g)T^2_A(g)=\frac12\cdot3^g\cdot(5^g-1),\\
\bar T^3(g)&=&\frac{1}{N(g)-3}T^3_{total}(g)=\frac{3^{g-1}(5^g-1)}{3^g-1}.
\end{eqnarray*}
Similarly, according to the symmetry of the Sierpinski Gasket network, the following relationship can be obtained:
\begin{eqnarray}
T^2_{total}(g)&=&\sum_{i\in\Omega(g)}T^2_i(g)
=\sum_{i\in\Omega(g)}T^3_i(g)+\big[\frac13(N(g)-3)+1\big]T^2_A(g)\nonumber\\
&=&T^3_{total}(g)+\frac13N(g)T^2_A(g)
=3^g\cdot5^g+\frac12(5^g-3^g).
\end{eqnarray}
Then, the mean time to absorptions on $SG(g)$ with double absorptive nodes satisfies:
\begin{eqnarray*}
\bar T^2(g)=\frac{1}{N(g)-2}T^2_{total}(g)=\frac{2\cdot3^g\cdot5^g+5^g-3^g}{3^{g+1}-1}.
\end{eqnarray*}

\bibliographystyle{unsrt}
\bibliography{Ref}

\end{document}